\documentclass[manuscript]{emulateapj}

\shorttitle{Molecular gas in young debris disks}

\shortauthors{Mo\'or et al.}

\begin{document}


\title{MOLECULAR GAS IN YOUNG DEBRIS DISKS}

\author{A. Mo\'or\altaffilmark{1}}
\email{moor@konkoly.hu}
\author{P. \'Abrah\'am\altaffilmark{1}}
\author{A. Juh\'asz\altaffilmark{2}}
\author{Cs. Kiss\altaffilmark{1}}
\author{I. Pascucci\altaffilmark{3}}
\author{\'A. K\'osp\'al\altaffilmark{2}}
\author{D. Apai\altaffilmark{3}}
\author{Th.~Henning\altaffilmark{4}}
\author{T. Csengeri\altaffilmark{5}}
\author{and C. Grady\altaffilmark{6,7}}  
\altaffiltext{1}{Konkoly Observatory of the Hungarian Academy of Sciences, P.O. Box 67, H-1525 Budapest, Hungary}
\altaffiltext{2}{Leiden Observatory, Leiden University, Niels Bohrweg 2, NL-2333 CA Leiden, The Netherlands}
\altaffiltext{3}{Department of Astronomy and Department of Planetary Sciences, The University of Arizona, Tucson, AZ 85721, USA}
\altaffiltext{4}{Max-Planck-Institut f\"ur Astronomie, K\"onigstuhl 17, 69117 Heidelberg, Germany}
\altaffiltext{5}{Max-Planck-Institut f\"ur Radioastronomie, Auf dem H\"ugel 69, 53121 Bonn, Germany}
\altaffiltext{6}{NASA Goddard Space Flight Center, Code 667, Greenbelt, MD 20771, USA}
\altaffiltext{7}{Eureka Scientific, 2452 Delmer Street, Suite 100, Oakland, CA 94602, USA}


\begin{abstract}

Gas-rich primordial disks and tenuous gas-poor debris disks are usually considered as two distinct 
evolutionary phases of the circumstellar matter. 
Interestingly, the debris disk around the young main-sequence star 49 Ceti
possesses a substantial amount of molecular gas, and possibly represents the missing link between 
the two phases. Motivated to understand the evolution of the gas component in circumstellar disks 
via finding more 49 Ceti-like systems, we carried out a CO $J$=3$-$2 survey
with Atacama Pathfinder EXperiment, targeting 20 infrared-luminous debris disks. These systems fill the 
gap between primordial and old tenuous debris disks in terms of fractional luminosity.
Here we report on the discovery of a second 49 Ceti-like disk around
the 30 Myr old A3-type star HD21997, a member of the Columba Association. 
This system was also detected in the CO(2$-$1) transition, and 
the reliable age determination makes it an even clearer example of an old gas-bearing 
disk than 49 Ceti.
While the fractional luminosities of HD21997 and 49 Ceti are not particularly high, 
these objects seem to harbor the most extended disks within our sample. 
The double-peaked profiles of HD21997 were reproduced 
by a Keplerian disk model combined with the LIME radiative 
transfer code.
Based on their similarities, 49 Ceti and HD21997 
may be the first representatives of a so far undefined new class of relatively old ($\gtrsim$8\,Myr), 
gaseous dust disks.  
From our results, neither primordial origin nor steady secondary production from icy planetesimals
can unequivocally explain the presence of CO gas in the disk of HD21997.

\end{abstract}





\section{Introduction} \label{intro}

Most young stars are surrounded by circumstellar disks, the natural by-product of star formation.
After a protostar has formed, its disk plays a crucial role in the evolution of the system, 
first by serving as reservoir for mass accretion, and later by becoming the birthplace of the planetary system.
At early evolutionary stage the mass of the {\sl primordial disk} is dominated by gas,
with a few percent of mass in small dust grains. The gaseous component
plays an important role in controlling the dust dynamics in the disk \citep{beckwith2000}, 
and its dispersal determines the time available to form Jupiter- and Saturn-like giant planets. 
As the disk evolves, the gas content is removed
by viscous accretion \citep{lynden1974} and/or by photoevaporation \citep[e.g.,][]{alexander2006}.

Current observational results 
imply that the primordial gas becomes largely depleted in the first 10\,Myr 
\citep{zuckerman1995,pascucci2006,fedele2010}.
A gas-poor disk appears, 
where the lifetime of individual dust grains under the influence of radiative forces -- without the stabilization effect of 
the surrounding gas -- 
is far less than the stellar age. 
The dust grains are believed to be continuously replenished 
by collisions and/or evaporation of previously formed planetesimals 
\citep[][and references therein]{wyatt2008}. 
In these {\sl debris disks} only a small amount of gas is expected.
Similarly to the dust, this gas could be {\sl second generation},
produced by sublimation of planetesimals, 
photodesorption from dust grains \citep{grigorieva2007}, or vaporization of colliding dust 
particles \citep{czechowski2007}.

So far only a few debris disks are known with detectable gas component.
The edge-on orientation of the disk around $\beta$\,Pic 
allowed the detection of a very small amount of circumstellar gas 
($N_{\rm CO}\sim6\times10^{14}$\,cm$^{-2}$)
through the presence of 
absorption/emission lines \citep[][]{roberge2000,brandeker2004}. \citet{redfield2007} successfully exploited the 
favorable edge-on geometry of the disk around \object[HD 32297]{HD32297} to detect gas via \ion{Na}{1}.
In contrast to the disks mentioned above, the debris disk around the young main-sequence star \object[49 Ceti]{49\,Ceti} 
seems to have substantial ($\sim$13\,$M_{\oplus}$) molecular gas \citep{zuckerman1995,dent2005,hughes2008}. 
The origin of the gas in the above mentioned systems is currently under debate. 
It can be residual primordial gas that survived longer in the outer disks than usually assumed  
\citep{krivov2009} or 
it may have formed or been released recently.
Based on dynamical arguments, \citet{fernandez2006} suggested that the gas in $\beta$\,Pic is secondary.
Analyzing high-resolution data obtained with the SMA interferometer at 230\,GHz, 
\citet{hughes2008} proposed that \object[49 Ceti]{49\,Ceti} is in a late stage of its evolution, possibly representing 
the link between gas-rich primordial disks and gas-poor optically thin 
debris disks. \citet{rhee2007} suggested an age of 20\,Myr for \object[49 Ceti]{49\,Ceti}, while \citet{thi2001} derived $\sim$8\,Myr as 
the age of the star. In the case of the older age, \object[49 Ceti]{49\,Ceti} would challenge the current picture 
of gas disk evolution, while the lower value could still be marginally consistent with a primordial disk phase.
The confirmation of the existence of debris disks containing a significant amount of gas would require to find and study 
more \object[49 Ceti]{49\,Ceti}-like systems with reliable age estimates.
It might well be possible that 
a number of similar systems exists, since most debris disks that are 
similar to \object[49 Ceti]{49\,Ceti} in terms of age and fractional 
luminosity have never been observed in molecular lines (most such candidates are in the southern hemisphere). 
Motivated by this fact, we carried out a survey with the APEX\footnote{This publication is based on data acquired with the Atacama
Pathfinder EXperiment (APEX). APEX is a collaboration between
the Max-Planck-Institut f\"ur Radioastronomie, the European Southern
Observatory, and the Onsala Space Observatory.} radio telescope 
to detect molecular gas in 20 infrared-luminous debris disks.
Here we review the results of this survey and report on the discovery of a second \object[49 Ceti]{49\,Ceti}-like disk around 
the 30\,Myr-old star \object[HD 21997]{HD21997}.

\section{Sample description} \label{sample}

For candidate selection we adopted the \object[49 Ceti]{49\,Ceti} system as a template. This A1-type star harbors a dusty 
disk that re-emits a fraction $f_{\rm dust}\sim8-9\times10^{-4}$ of the star's radiation at infrared 
wavelengths. This fractional luminosity is an order of magnitude lower than the corresponding value
in primordial disks. 
We used the following target selection criteria: 
(1) spectral type between A0 and F6;
(2) $f_{\rm dust}$ in the range of 
$ 5\times 10^{-4} to 5\times 10^{-3}$ that excludes both primordial disks and very tenuous debris disks; 
(3) the excess emission is confirmed by an instrument independent of {\sl IRAS} 
(4) ages between 10 and 100\,Myr, the age estimate is based on kinematic group membership or other 
reliable dating methods.
In total, 20 candidates were selected from the lists of \citet{chen2005} and \citet{moor2006,moor2010}. 
Their basic properties are given in Table~\ref{table1}.

\begin{figure} 
\epsscale{0.86}
\plotone{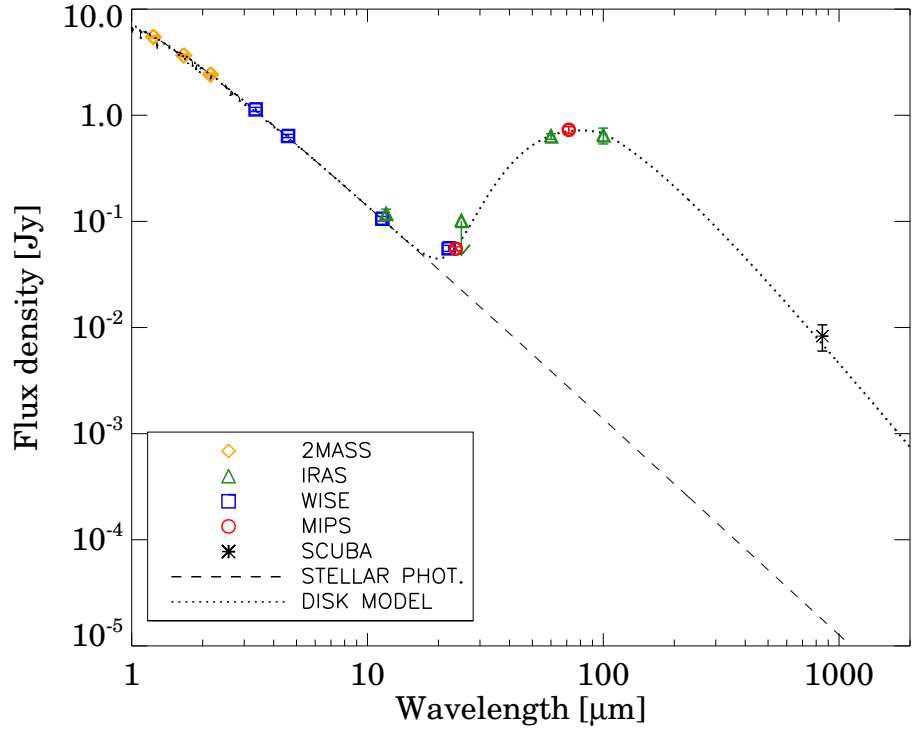}
\caption{{\sl Spectral energy distribution of HD21997. Photometric data presented in this figure are color-corrected.}
 \label{fig1}}
\end{figure}

For most sources, disk parameters (temperature, radius, and 
fractional luminosity) were adopted from the literature (Table~\ref{table1}).
In all cases, disk radii were estimated by 
adopting blackbody-like dust emission. For \object[HD 121617]{HD121617} no previous literature 
data were found, thus we collected infrared photometry from the {\sl IRAS} FSC, 
{\sl AKARI} IRC, {\sl AKARI} FIS and {\sl WISE} \citep{wright2010}. For \object[HD 21997]{HD21997}, we also compiled a new 
spectral energy distribution 
based on literature data and our own photometry derived from archival observations obtained with the 
Multiband Imaging Photometer for Spitzer \citep[see Figure~\ref{fig1}, 55.3$\pm$2.2\,mJy at 24\,{\micron} 
 and 662$\pm$47\,mJy at 70\,{\micron}; for the description of the processing see ][]{moor2010}. 
Disk parameters for these two targets were derived following \citet{moor2010}.
 For sources where submillimeter observations were available,
we computed dust masses assuming optically thin emission with $\kappa_{870\micron}=2$\,cm$^2$\,g$^{-1}$ and 
$\beta = 1$ \citep{nilsson2010}, and dust temperature from Table~\ref{table1}.
For comparison of the fundamental parameters, \object[49 Ceti]{49\,Ceti} is also added to Table~\ref{table1}.

\section{Observations and data reduction} \label{obsanddatared}

Our survey was carried out 
with the APEX 12\,m telescope \citep{gusten2006} in service mode, between 2008 October and 2009 November.
All objects were observed at the 345.796\,GHz $^{12}$CO $J$=3$-$2 line using the 
SHeFI/APEX2 receiver 
\citep{vassilev2008}.
One source, \object[HD 21997]{HD21997} was also 
observed in the $J$=2$-$1 transition of $^{12}$CO (at 230.538\,GHz) with the SHeFI/APEX1 receiver.
For the backend, we used the
Fast Fourier Transform Spectrometer with 2048 channels providing a velocity resolution of 0.42 and 0.64\,kms$^{-1}$ in the 
$J$=3$-$2 and $J$=2$-$1 transitions, respectively. An on-off observing pattern was utilized with beam switching.
The total on-source integration time for most sources ranged between 10 and 30\,minutes. For \object[HD 21997]{HD21997}, 
we integrated 
longer for a better characterization of the line profile. 
The weather conditions were generally dry,
the precipitable water vapor was below 1.3\,mm for most of the $J$=3$-$2 observations and ranged between 
1.3 and 2.7\,mm during the $J$=2$-$1 measurements. 
The data reduction was performed using the GILDAS/CLASS package\footnote{\url{http://iram.fr/IRAMFR/GILDAS/}}. 
For the final average spectrum, we omitted {noisy scans} 
and a linear baseline was subtracted from each individual scan. 

\section{Results and analysis} \label{analysis}

\begin{figure} 
\epsscale{0.86}
\plotone{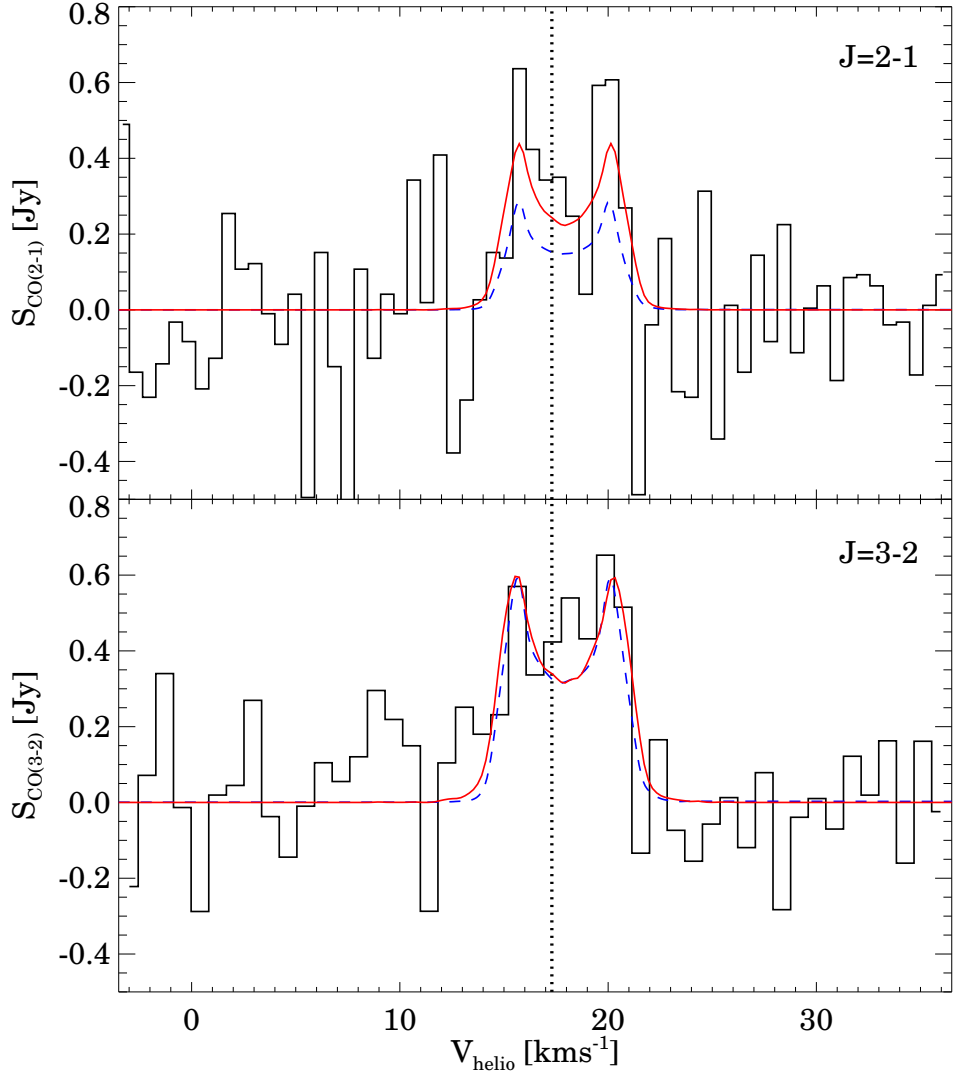}
\caption{{\sl CO 
spectra of HD21997. The CO(3$-$2) spectrum is binned by a factor of two.
Dotted line marks the radial velocity of the star. 
The dashed line (blue in online version) corresponds to a disk model with H$_2$/CO=10$^4$, the solid line 
(red in online version) to H$_2$/CO=10$^3$.
}
 \label{fig2}}
\end{figure}

Among the 20 targets, one system, \object[HD 21997]{HD21997}, was detected at $>$5\,$\sigma$ level in both CO lines. 
Figure~\ref{fig2} shows the baseline-corrected CO profiles for \object[HD 21997]{HD21997}. 
Both lines display double peaked line profile with identical peak positions.
The central velocities of both lines are consistent with the systemic velocity of the star \citep[+17.3$\pm$0.8kms$^{-1}$;][]{kharchenko2007}. 
We integrated the intensities/fluxes over an interval of 
8\,km\,s$^{-1}$ that covers the whole line profile.
The beam efficiencies and Kelvin-to-Jansky conversion factors were taken from the APEX web
page\footnote{\url{http://www.apex-telescope.org/telescope/efficiency/}}. 
For the non-detected sources, upper limits were estimated 
as $T_{\rm rms} \Delta{v} \sqrt{N}$, where $T_{\rm rms}$ is the measured noise, $\Delta{v}$ is the velocity channel width, 
and $N$ is the number of velocity channels over an interval of 10\,km\,s$^{-1}$. 
The total mass (or upper limit) of CO molecules ($M_{\rm CO}$) was estimated assuming optically thin emission and local thermodynamic equilibrium (LTE). 
The excitation temperature ($T_{\rm ex}$) was assumed to be equal to the dust temperature in Table~\ref{table1} (i.e., 
gas and dust are sufficiently coupled).
The obtained line intensities/fluxes as well as the estimated CO masses are listed in Table~\ref{table1}.
Note that for \object[HD 21997]{HD21997} the CO masses computed independently from the (2$-$1) and (3$-$2) transitions 
are significantly different.

\begin{figure*} 
\epsscale{1.1}
\plotone{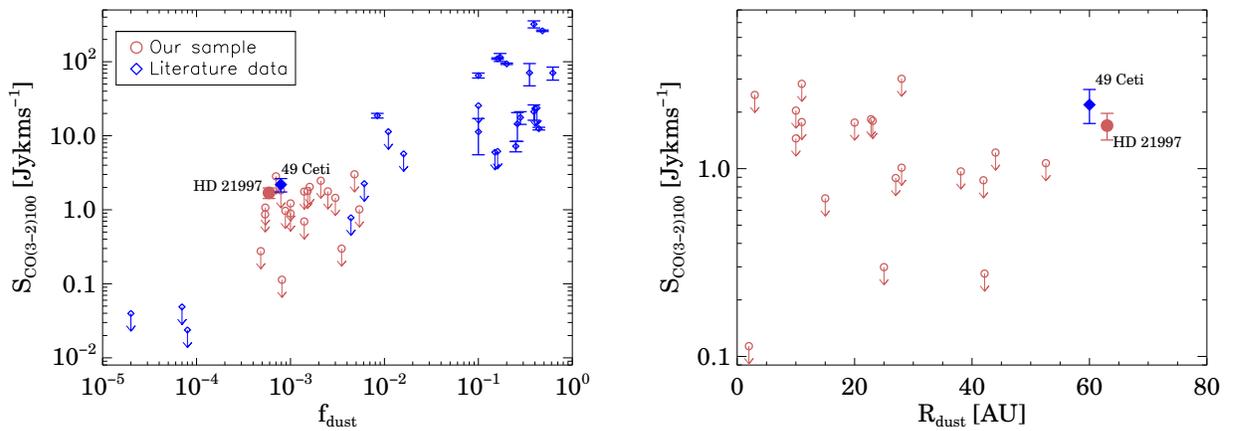}
\caption{ {\sl Left panel: integrated CO(3$-$2) fluxes or upper limits
for our sample, 
and for protoplanetary/debris disks around A0-F6-type stars 
from literature data \citep{dent2005,greaves2000a,greaves2000} 
normalized to 100\,pc are plotted against fractional luminosities. 
49\,Ceti \citep{dent2005}, and HD21997 
are plotted with larger filled symbols. 
Right panel: integrated CO(3$-$2) fluxes for 
our sample and for 49\,Ceti as a function of disk radii.}
\label{fig3}}
\end{figure*}

Figure~\ref{fig3}(left panel) shows the integrated CO(3$-$2) fluxes and upper limits,
normalized to 100\,pc, plotted against the fractional luminosities of the disks.
For comparison, additional protoplanetary/debris disks around A0-F6-type pre-main/main-sequence stars, including \object[49 Ceti]{49\,Ceti},  
are also displayed \citep{dent2005,greaves2000a,greaves2000}. 
Our observations fill the gap between Herbig Ae/Be disks and older debris disks.
Note that the fractional luminosities of \object[HD 21997]{HD21997} and  
 \object[49 Ceti]{49\,Ceti} are modest even within the debris sample, and significantly lower than those of the primordial disks.
Thus, $f_{\rm dust}$ does not appear to be a good proxy for the presence of CO gas in debris disks. 
Figure~\ref{fig3}(right panel) presents 
the integrated CO(3$-$2) fluxes versus 
the (blackbody) radii of the dust disks. Interestingly, the two definite detections, \object[HD 21997]{HD21997} 
and \object[49 Ceti]{49\,Ceti}, 
harbor the most extended disks, suggesting that large radius and low dust temperature may be essential for CO detection.
Although the radii in this analysis rely on the assumption of blackbody grains, the conclusion holds in the case of realistic 
grain size distributions. Using the blackbody assumption, we may systematically underestimate the true radius, due to 
the presence of inefficiently emitting small particles. However, assuming similar 
grains in all disks, the relative distribution of disk radii would not differ from the blackbody case 
\citep[see e.g.,][]{wyatt2008}.

\object[HD 21997]{HD21997} is an A3-type star 
at a distance of 72\,pc \citep{vanleeuwen07}, a member of the $\sim$30\,Myr old Columba group \citep{moor2006,torres2008}. 
Fitting an ATLAS9 atmosphere model \citep{castelli2003} to the optical and near-IR ({\sl Hipparcos, Tycho-2, Two Micron All Sky Survey}) data, 
assuming solar metallicity and $\log{g} = 4.25$ yields $T_{\rm eff}$ = 8300\,K. The evolutionary tracks of \citet{siess2000} 
imply a stellar mass of 1.85\,$M_{\sun}$.
We modeled the measured line profiles of \object[HD 21997]{HD21997} with a simple disk geometry assuming 
a combination of a radial power-law and a vertical Gaussian profile for the density distribution: 
\begin{equation}
n_{\rm CO}(r,z)=n_{\rm CO,in}(r/R_{\rm in})^{-\alpha}e^{-z^2/2H^2}. 
\end{equation}
We fixed the following parameters: $H=0.1r$, $\alpha=-2.5$, $R_{\rm in}=63$\,AU (Table~\ref{table1}), and
$R_{\rm out}=200$\,AU \citep[typical for Herbig Ae/Be disks;][]{panic2009}. 
We assumed an H$_2$ abundance relative to CO of 10$^{4}$, and that gas and dust grains -- the latter act like blackbody -- are well mixed, 
prescribing that the gas kinetic temperature and dust temperature distributions are identical.
The velocity of the material in the disk was derived by assuming
Keplerian orbits around a star of 1.85\,$M_\sun$ mass.
Then the CO level populations at each disk position, and the resulting emission line 
profiles were calculated using the non-LTE spectral line radiation transfer code LIME \citep{brinch2010}. 
First, we fitted the (3--2) line by adjusting $n_{\rm CO,in}$ and the disk inclination.
The best-fitting model spectrum with $n_{\rm CO,in}=10$\,cm$^{-3}$ and $i=45^\circ$ is overplotted with dashed line 
in Figure~\ref{fig2}.
With the same parameters we also computed a CO (2--1) profile. As Figure~\ref{fig2} shows, 
this model significantly underestimates the observed CO(2--1) feature.
The reason for this discrepancy
is the same as what causes the difference in the CO mass estimates from (2--1) and (3--2) 
lines:  the ratio of integrated CO(3$-$2) flux to the
integrated CO(2$-$1) flux is only 1.43$\pm$0.37, significantly 
lower than the ratio of 3.8, expected for $T_{\rm ex}\sim60$\,K in LTE condition.
 This low line flux ratio corresponds to an excitation temperature of 
13.1$\pm$2.7\,K. This is unrealistically low (subthermal) for being an LTE value,
suggesting that the density of collision partners (H$_2$) 
is lower than the critical density, and the excitation
is not collisionally dominated.
In order to provide a model that can fit both lines simultaneously, we gradually decreased the  
H$_2$/CO abundance ratio and repeated the above-mentioned modeling process. 
We found that with H$_2$/CO=1000$\pm$500, $n_{\rm CO,in}=22\pm5$\,cm$^{-3}$, and $i=45^{+15}_{-10}$, both line profiles 
can be fitted (solid line in Figure~\ref{fig2}). Note that in this non-LTE model the 
kinetic and excitation temperatures of the gas are different.
The total CO mass predicted by this model is $M_{\rm CO}=3.5\times10^{-4}$\,$M_{\oplus}$.


\section{Discussion} \label{discussion}

Its reliable age determination makes \object[HD 21997]{HD21997} the oldest example for a gas-bearing debris disk.
In many aspects it resembles the somewhat younger \object[49 Ceti]{49\,Ceti}.
Both systems contain an A-type central star that produces energetic UV that could dissociate CO molecules in the vicinity of the star.
\object[49 Ceti]{49\,Ceti} and \object[HD 21997]{HD21997} clearly stand out from our sample in terms of 
disk radius, and of harboring a large amount of relatively cold dust ($T_{\rm dust}\leq80$\,K and $M_{\rm dust}\sim0.1$\,$M_{\oplus}$). 
{Note also that these two systems exhibit very similar $M_{\rm CO}$/$M_{\rm dust}$ ratios ($\sim$0.003).} 
Based on their similarities, 
we speculate that \object[HD 21997]{HD21997} and \object[49 Ceti]{49\,Ceti}
may be the first representatives of a so far 
undefined new class of relatively old ($\gtrsim$8\,Myr), gaseous dust disks.

\citet{hughes2008} claim that the disk around \object[49 Ceti]{49\,Ceti} contains predominantly primordial gas 
and may represent the link between gas-rich primordial disks and gas-poor debris disks. 
It is a question whether \object[HD 21997]{HD21997} system may be of similar origin.
The gas clearing process in primordial disks is expected to progress outwards 
due to photoevaporation driven by the central star
\citep[e.g.,][]{alexander2006,pascucci2010},
 thus the last reservoir of gas will be the outermost part of the circumstellar disk. 
Indeed, \object[HD 21997]{HD21997} and \object[49 Ceti]{49\,Ceti} possess the largest disks and consequently have the longest expected 
survival time for gas in our sample (though the evaporation timescale also depends on the high energy flux 
of the central star). A confirmed primordial origin of the gas in the \object[HD 21997]{HD21997} system, 
would pose a serious question to the current paradigm 
since its age of $\sim$30\,Myr significantly exceeds both the model predictions 
for disk clearing and the ages of the oldest T\,Tauri-like or transitional gas disks in the literature \citep{kastner2008}.

Primordial CO gas can survive only in the case of efficient shielding from the stellar/interstellar high-energy photons.
We determined the stellar UV flux from the fitted ATLAS9 atmosphere model (see Section~\ref{analysis}) and the UV component of 
the interstellar radiation field (ISRF) from \citet{habing1968}.
For each disk position, where the H$_2$ and CO column densities are provided by our model, we 
analyzed the shielding efficiency using the photodissociation model of \citet{visser2009} 
for different $N_{\rm CO}$ and $N_{\rm H_2}$ pairs (shielding by dust grains is negligible in this tenuous disk). 
We found that no region in the disk is shielded enough to provide a CO lifetime longer than 500\,yr. 
Adopting a lower scale height would lead to higher radial column densities but would not 
affect significantly the vertical column densities, thus the UV photons of the ISRF -- which dominate the stellar UV flux 
almost everywhere in the disk -- could efficiently photodissociate CO molecules. 
In the course of modeling we assumed that the gas and dust are sufficiently coupled leading to a common
temperature, but in tenuous debris disks this may not be true \citep{kamp2001,zagorovsky2010}. 
Assuming a lower gas temperature similar to the measured excitation temperature would allow the existence of a larger amount of 
hydrogen gas. However, the gas is not likely to cool down to such a low temperature all over the disk. 
{Thus based on this result, as well as on the obtained H$_2$/CO ratio of $\sim$1000 that is lower 
than expected for a primordial composition, 
the scenario of primordial origin of gas in \object[HD 21997]{HD21997} is unlikely.}

Is it possible then, that the gas in this disk is of secondary origin, being continuously replenished from icy planetesimals? 
In this scenario the gas may have a very low H$_2$/CO ratio.
Without the presence of a large amount of H$_2$, 
shielding against UV photons is weak. 
Our modeling predicts CO photodissociation timescales of less than 500 years in the whole disk. 
In order to reproduce the observed CO mass of $\sim$3.5$\times10^{-4}$\,{$M_\oplus$}, CO has to be released from solids with a rate of 
$>$7$\times10^{-7}$\,{$M_\oplus$}yr$^{-1}$. 
Pure CO ice evaporates at temperatures above 18\,K, thus it is volatile
even far from the luminous central star, meaning that the surface of planetesimals is very likely already depleted 
in CO ice. However, CO ice can persist in deeper layers and/or can be trapped in mixed H$_2$O--CO ices even on the 
surface at temperatures below $\sim$140\,K. 
Destructive collisions between planetesimals can lead to the release of subsurface ices.
In addition, frequent collisions of icy grains with smaller particles --  mainly with $\beta$ meteoroids -- 
could produce a continuous flux of CO via vaporization. Photodesorption from solids can also 
significantly contribute to the gas production.
Extrapolation of the current production rate for the last 20\,Myr (assuming a primordial gas-rich disk phase in the first 10\,Myr 
and a steady-state disk evolution afterwards)
would yield a total of $>$14\,{$M_\oplus$} of CO released from 
planetesimals/grains. Adopting a CO mass abundance of 10\% in the planetesimals 
\citep[see the composition of the comet Hale-Bopp;][]{huebner1999},
this scenario would require the complete destruction of more than 140\,{$M_\oplus$} mass of planetesimals in the outer disk. 
It would significantly exceed 
the full initial dust content of a typical protoplanetary disk, making this steady-state scenario questionable. 
A more satisfactory explanation would be that the system is currently undergoing a phase of temporarily high CO production.
The origin of this contemporeous gas production might be imprinted in the spatial distribution 
of the gas and dust, and could be revealed with future interferometers.

Our results indicate that neither primordial origin nor steady secondary production
can unequivocally explain the presence of CO gas in the disk of \object[HD 21997]{HD21997}. 
An on-going temporarily high CO production may be more likely. 
Detection of other gas components and transitions with {\sl Herschel} and {\sl ALMA}, 
as well as the better characterization of the disk structure may lead to the deeper understanding 
of this enigmatic system and clarify whether \object[49 Ceti]{49\,Ceti} and \object[HD 21997]{HD21997}
are the first examples of a so far less studied phase of disk evolution.

\acknowledgments
We thank an anonymous referee whose comments significantly improved the manuscript.
We are grateful to the APEX staff, in particular to Carlos De Breuck (ESO), for their assistance.
This research was partly funded by the Hungarian OTKA grant K81966. The research of
\'A.K. is supported by the Netherlands Organisation for Scientific Research.

{\it Facilities:} \facility{APEX}.

\clearpage

\tabletypesize{\tiny}
\begin{deluxetable}{lccccccccccccccc}
\tabletypesize{\tiny}                                                                                                                           
\tablecaption{Stellar/disk properties. \label{table1}}                                                                                                
\tablewidth{0pt}                                                                                                                                      
\tablecolumns{16}                                                                                                                                     
\tablehead{ 
\colhead{(1)} & \colhead{(2)} & \colhead{(3)} &                                                                                                          
 \colhead{(4)} & \colhead{(5)} & \colhead{(6)} &  \colhead{(7)}                                                                              
& \colhead{(8)} & \colhead{(9)} & \colhead{(10)} & \colhead{(11)} &                                                                  
\colhead{(12)} &\colhead{(13)} & \colhead{(14)} & \colhead{(15)} & \colhead{(16)}\\ 
\colhead{ID} & \colhead{Sel.} & \colhead{SpT} &  \colhead{D} & \colhead{Memb.} & \colhead{v$_{\rm rad}$} &                                
\colhead{$T_{\rm dust}$} &  \colhead{$R_{\rm dust}$} &  \colhead{$f_{\rm dust}$} &  \colhead{Ref.} & \colhead{$M_{\rm dust}$} & \colhead{Ref.} &      
\colhead{t$_{\rm on}$} & \colhead{I$_{\rm CO}$} & \colhead{S$_{\rm CO}$} & \colhead{$M_{\rm CO}$} \\                                                      
\colhead{} & \colhead{} & \colhead{} & \colhead{(pc)} & \colhead{} & \colhead{(km\,s$^{-1}$)} &                                                       
\colhead{(K)} & \colhead{(AU)} & \colhead{(10$^{-4}$)} & \colhead{} & \colhead{($M_{\earth}$)} & \colhead{} &                                     
\colhead{(minute)} &\colhead{(K\,km\,s$^{-1}$)} &  \colhead{(Jy\,km\,s$^{-1}$)} & \colhead{($\rm 10^{-4} M_{\earth}$)}                                     
}                                                                                                                                                     
\startdata                                                                                                                                            
           HD 3670 &   1 &        F5V &     (76) &    ColA &    $+$8.6 &   53 &   41 &    5.4 &    1 &    \nodata &      &    35.5 &                  $<$0.050 &                   $<$1.50 &                   $<$0.85 \\
          HD 15115 &   1 &         F2 &      45  &    BPMG &    $+$8.8 &   56 &   42 &    4.8 &    1 &      0.036 &    1 &    39.4 &                  $<$0.045 &                   $<$1.35 &                   $<$0.28 \\
          HD 21997 &   2 &     A3IV/V &      72  &    ColA &   $+$17.3 &   64 &   63 &    5.9 &    5 &      0.14  &    1 &   255.3 &  0.109$\pm$0.014 &             {3.28$\pm$0.53} &      1.79$\pm$0.32 \\
          HD 21997\tablenotemark{a} &     &            &          &         &           &      &      &           &      &         &      &   115.9 &    {0.078$\pm$0.014} &       {2.29$\pm$0.47} &         4.9$\pm$1.1 \\
          HD 30447\tablenotemark{b} &   1 &        F3V &      80  &    ColA &   $+$21.3 &   62 &   38 &    8.8 &    1 &       0.16 &    2 &    24.8 &                  $<$0.050 &                   $<$1.50 &                   $<$1.01 \\
          HD 32297\tablenotemark{b} &   2 &        A0V &     112  &    \nodata &   $+$23.0 &   85 &   28 &   54.0 &    2 &       0.55 &    3 &    73.6 &                  $<$0.027 &                   $<$0.80 &                   $<$1.25 \\
          HD 35841 &   1 &        F5V &     (96) &    ColA &   $+$23.1 &   68 &   23 &   15.2 &    1 &    \nodata &      &    13.5 &                  $<$0.065 &                   $<$1.95 &                   $<$1.96 \\
          HD 38207 &   2 &        F2V &     (93) &    ColA &   $+$24.9 &   59 &   44\tablenotemark{c} &   10.0    &    3 &    $<$0.28 &    4 &    22.4 &                  $<$0.047 &                   $<$1.41 &                   $<$1.24 \\
         HD 106906 &   3 &        F5V &      92  &     LCC &   $+$11.1 &   90 &   20 &   14.0 &    4 &    \nodata &      &    24.2 &                  $<$0.069 &                   $<$2.08 &                   $<$2.24 \\
         HD 110058 &   2 &        A0V &     107  &     LCC &   $+$5.0  &  130 &   11 &   25.0 &    2 &    $<$0.34 &    5 &    43.0 &                  $<$0.051 &                   $<$1.54 &                   $<$2.90 \\
         HD 113766 &   3 &        F4V &     123  &     LCC &    $-$0.6 &  330 &    3 &   21.0 &    4 &    \nodata &      &    28.0 &                  $<$0.055 &                   $<$1.64 &                   $<$8.75 \\
         HD 114082 &   3 &        F3V &      85  &     LCC &    $+$5.2 &  110 &   10 &   30.0 &    4 &    \nodata &      &    20.6 &                  $<$0.066 &                   $<$1.98 &                   $<$2.11 \\
         HD 115600 &   3 &     F2IV/V &     110  &     LCC &    $+$4.7 &  120 &   10 &   16.0 &    4 &    \nodata &      &    24.0 &                  $<$0.056 &                   $<$1.67 &                   $<$3.15 \\
         HD 117214 &   3 &        F6V &     110  &     LCC &    $+$7.2 &  110 &   11\tablenotemark{c} &    7.0 &    4 &    \nodata &      &    10.8 &                  $<$0.078 &                   $<$2.33 &                   $<$4.12 \\
         HD 121617 &   2 &        A1V &    (120) &     UCL &   $+$13.6 &  105 &   28 &   48.0 &    5 &    $<$0.33 &    6 &    14.2 &                  $<$0.070 &                   $<$2.09 &                   $<$4.24 \\
         HD 164249 &   2 &        F5V &      48  &    BPMG &    $-$0.2 &   70 &   27 &   10.0 &    2 &   $<$0.076 &    7 &     6.7 &                  $<$0.129 &                   $<$3.85 &                   $<$0.99 \\
         HD 172555 &   2 &        A7V &      29  &    BPMG &    $+$2.0 &  320 &    2 &    8.1 &    2 &    \nodata &      &    26.2 &                  $<$0.047 &                   $<$1.39 &                   $<$0.39 \\
         HD 181327 &   2 &        F6V &      52  &    BPMG &    $+$0.2 &   75 &   25 &   35.0 &    2 &       0.40 &    7 &    31.6 &                  $<$0.037 &                   $<$1.11 &                   $<$0.34 \\
         HD 191089\tablenotemark{b} &   2 &        F5V &      52  &    BPMG &    $-$5.8 &   95 &   15 &   14.0 &    2 &      0.022 &    4 &    17.1 &                  $<$0.085 &                   $<$2.55 &                   $<$0.92 \\
         HD 192758 &   1 &        F0V &     (62) &   Argus &   $-$11.1 &   56 &   52 &    5.4 &    1 &    \nodata &      &     9.8 &                  $<$0.093 &                   $<$2.78 &                   $<$1.07 \\
         HD 221853\tablenotemark{b} &   1 &         F0 &      68  &      LA &    $-$4.2 &   83 &   22 &    7.9 &    1 &    \nodata &      &     3.4 &                  $<$0.131 &                   $<$3.91 &                   $<$2.24 \\
\tableline
          49\,Ceti &     &        A1V &      59  &   \nodata &           &   80 &   60 &    7.9 &    2 &      0.074 &    8 &         &                        	    &                           &             2.60$\pm$0.54\tablenotemark{d} \\
  \enddata                                                                                                                                              
 \tablecomments{Column\,1: identification.                                                                                                            
 Column\,2: reference for the selection. 1--\citet{moor2010}, 2--\citet{moor2006},                                                                  
 3--\citet{chen2005}.                                                                                                                                
 Column\,3: spectral type.                                                                                                                            
 Column\,4: distance.                                                                                                                                 
 Parenthesis indicate photometric or kinematic distances, otherwise Hipparcos distances from \citet{vanleeuwen07} are used.                           
 Column\,5: membership status. ColA: Columba Association; BPMG: $\beta$\,Pic moving group, LCC:                                           
 Lower Centaurus Crux association, UCL: Upper Centaurus Lupus association, Argus: Argus moving group,                                                 
 LA: Local Association.                                                                                                                               
 Column\,6: heliocentric radial velocity.                                                                                                              
 Column\,7: disk temperature.                                                                                                                         
 Column\,8: disk radius.                                                                                                             
 Column\,9: fractional dust luminosity. 
 Column\,10: references for disk parameters. (1) \citet{moor2010}, (2) \citet{rhee2007}, (3) \citet{hillenbrand2008},                                 
 (4) \citet{chen2005}, (5) this work.                                                                                                                 
 Column\,11: dust mass.                                                                                                                               
 Column\,12: reference for submillimeter measurement that was used in the dust mass estimate. (1) \citet{williams2006},                               
 (2) \citet{nilsson2010}, (3) \citet{maness2008}, (4) \citet{roccatagliata2009}, (5) \citet{sylvester2001},                                           
 (6) \citet{sheret2004}, (7) \citet{nilsson2009}, (8) \citet{song2004}.                                                                
 Column\,13: on-source integration time.                                                                                                              
 Column\,14: line intensity for CO $J$=3$-$2. Intensity units are main-beam brightness                                                                   
 temperature.                                                      
 Column\,15: integrated line flux.                                                                    
 Column\,16: estimated mass of the CO gas.
 } 
\tablenotetext{a}{Parameters from the $J$=2$-$1 line.}
\tablenotetext{b}{\citet{kastner2010} reported the nondetection of CO(2$-$1) emission.}
\tablenotetext{c}{For those objects where our distance estimate differs from the literature value by $>$10\%, 
the R$_{\rm dust}$ was rescaled according to our distance.}
\tablenotetext{d}{CO mass of 49\,Ceti was derived using the CO(3--2) observation of \citet{dent2005}.}                                                                                                                                                    
\end{deluxetable}

\clearpage


\end{document}